%% file: IC23_onlineSD_draft.tex
\documentclass{article}
\usepackage{spconf,graphicx}
\usepackage{amsmath,amssymb}
\usepackage{url,lipsum,cite}
\usepackage{multicol,multirow}
\usepackage{xpatch,xcolor}
\usepackage{tabularx}
\usepackage{makecell, booktabs}
\usepackage{algorithm}
\usepackage{algpseudocode}
\usepackage{slashbox}
\newcolumntype{Y}{>{\centering\arraybackslash}X}
\usepackage{hyperref}

\hypersetup{
    colorlinks=true,
    linkcolor=purple,
    filecolor=magenta,      
    urlcolor=purple,
}

\newcommand{\newpara}[1]{\vspace{8pt}\noindent\textbf{#1}}
\def\BigRoman{\uppercase\expandafter{\romannumeral\number\count 255 }}
\def\Romannumeral{\afterassignment\BigRoman\count255=}

\title{Absolute decision corrupts absolutely: conservative online\\speaker diarisation}

\name{Youngki Kwon, Hee-Soo Heo, Bong-Jin Lee, You Jin Kim, and Jee-weon Jung}
\address{Naver Corporation, South Korea}
\begin{document}
\ninept
\maketitle
\begin{abstract}
Our focus lies in developing an online speaker diarisation framework which demonstrates robust performance across diverse domains.
In online speaker diarisation, outputs generated in real-time are irreversible, and a few misjudgements in the early phase of an input session can lead to catastrophic results. 
We hypothesise that cautiously increasing the number of estimated speakers is of paramount importance among many other factors.
Thus, our proposed framework includes decreasing the number of speakers by one when the system judges that an increase in the past was faulty.
We also adopt dual buffers, checkpoints and centroids, where checkpoints are combined with silhouette coefficients to estimate the number of speakers and centroids represent speakers. 
Again, we believe that more than one centroid can be generated from one speaker. 
Thus we design a clustering-based label matching technique to assign labels in real-time. 
The resulting system is lightweight yet surprisingly effective. 
The system demonstrates state-of-the-art performance on DIHARD {\Romannumeral 2} and {\Romannumeral 3} datasets, where it is also competitive in AMI and VoxConverse test sets. 
\end{abstract}

\section{Introduction}
\label{sec:intro}
Speaker diarisation (SD), which segments input audio to short utterances according to speaker identity, is going through a rapid breakthrough~\cite{anguera2012speaker,park2022review}.
Based on the success of recent SD systems~\cite{Fujita2019,fujita2019end,kinoshita2021integrating,kinoshita2022tight,Horiguchi2020,medennikov2020target,landini2022bayesian,kwon2021adapting,kwon2021look,wang2018speaker}, online SD systems are also being developed~\cite{soldi2015adaptive,zhu2016online,zhang2022low,yue2022online,wang22j_interspeech,han2021bw,xue2021online,fini2020supervised}.
In an online SD system, the system should decide the speaker label of a given short segment leveraging only current and past segments, where only a part of past segments are available.
Since the system cannot observe the whole audio in each decision step, speaker diarisation performance tends to be worse than the offline SD systems. 
Moreover, those systems require the characteristic of real-time, so the online system's components should be light-weighted. 

Early works on online SD adopted a universal background model or an i-vector with log-likelihood or cosine similarity-based clustering~\cite{soldi2015adaptive,zhu2016online}.
With advances in deep learning, more recent works rely upon either speaker embeddings extracted from a deep neural network (DNN) or entirely compose an online end-to-end neural diarisation (EEND) models~\cite{xue2021online,horiguchi2022online,han2021bw}.
Xue et al.~\cite{xue2021online} introduced EEND-based online diarisation systems using speaker-tracing buffer (STB). Han et al.~\cite{han2021bw} presented the variants of the EEND-EDA system working as a block-wise online inference.
In \cite{zhang2022low}, authors modified the agglomerative hierarchical clustering (AHC) algorithm, widely adopted in offline SD systems, and proposed a checkpoint AHC with the label matching algorithm.
Authors of \cite{yue2022online} adopted a memory module for each speaker and contained selected embeddings, where VBx~\cite{landini2022bayesian} and cosine operations on centroids were used for clustering. 
Wang et al.~\cite{wang22j_interspeech} adapted target speaker voice activity detection (TS-VAD), a successful offline SD framework, to online SD scenarios~\cite{medennikov2020stc, medennikov2020target}.
As mentioned above, the literature is witnessing diverse frameworks. 

The checkpoint AHC system~\cite{zhang2022low} stores the intermediate result of clustering as a checkpoint so that AHC can meet real-time operation requirements.
Here, the checkpoint consists of embeddings, where hundreds of embeddings are employed to represent all past sequences compactly.
Whenever a new embedding arrives, all embeddings stored so far are clustered, and a label is assigned to the most recent embedding via a bipartite label matching algorithm.  
When the embeddings outnumber the checkpoint size, the continuously updated checkpoint is stored instead of all embeddings. 
In order to keep the size of the checkpoint, the two most similar embeddings in the checkpoint are combined.

We analyse that there are several essential components which an online SD should include compared to an offline SD system. 
First, a fixed-size buffer is required to store embeddings because entire embeddings cannot be clustered concurrently.
The buffer can exist in various forms where the size is identical to the number of a detected speaker at least (e.g., one centroid for each speaker) or much larger (e.g. hundreds of checkpoint embeddings).
Second, precisely tracking the number of speakers detected so far is required. 
This idea includes judging whether an embedding input to the system every step is from an unseen speaker or one of the existing speakers.
Third, a method for solving the permutation between each clustering is required for systems that conduct local clustering multiple times with one or a few new embeddings. 
\begin{figure*}[ht!]
    \centering
    \includegraphics[width=\textwidth]{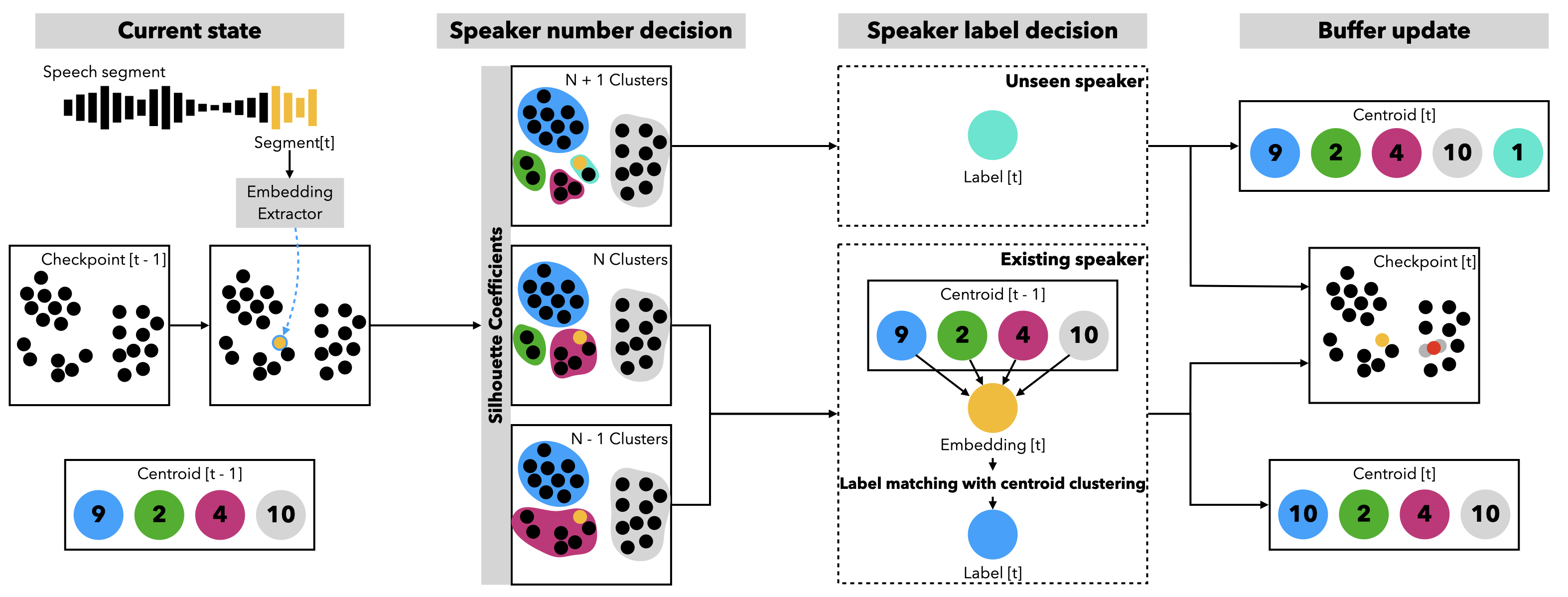}
    \caption{
      The overall scheme of the proposed online speaker diarisation system.
      }
    \vspace{-5pt}
    \label{fig:framework}
\end{figure*}

In this work, we propose a novel online SD framework.
The framework is built upon several recent frameworks as well as novel components that we introduce. 
We make three main proposals. 
First, we adopt silhouette coefficient~\cite{rousseeuw1987silhouettes} for the first time in online SD. 
It is used for two purposes: (i) judges whether an input embedding is from an unseen speaker and (ii) initially estimates the number of speakers at the current time. 
Second, we adopt a dual buffer system where we leverage both the checkpoint proposed in \cite{zhang2022low} and centroids which serve as the `{\em candidate}' speakers\footnote{We refer to the centroid as a candidate because there can be situations where more than one centroid is generated for a single speaker.}.
Third, we propose to stack input embeddings up to a specific moment ($30$ seconds in our case) where we perform offline AHC after that moment.
Then, our system starts to operate as an online SD system. 
The main philosophy behind our proposed online SD framework is that miscellaneous errors inevitably exist because it is an ``{\em online}'' system. 
This philosophy affected several design choices, including stacking embeddings up to a certain timestep, enabling a decrease in the estimation of speaker numbers, and clustering centroids once again. 

We show that with silhouette coefficients, we can estimate the number of speakers both in an unsupervised manner and in real-time.
The dual buffer clustering technique greatly improves the performance. 
In our analysis, this counteracts a single speaker represented by multiple centroids.
Having a short amount of time before operating as an online SD stabilises the system.
Combining these proposals, we demonstrate that our proposed online SD system achieves the new state-of-the-art performance on DIHARD {\Romannumeral 2} and {\Romannumeral 3} evaluation sets, which is the most widely adopted evaluation scenario many preceding works have investigated.

The rest of this paper is organised as follows.
Section~\ref{sec:proposed} introduces our proposed online SD framework in detail.
Section~\ref{sec:exp} and \ref{sec:result} addresses experiments and corresponding results. 
The paper is concluded, and future directions are given in Section~\ref{sec:conclusion}.

\section{Proposed online SD pipeline}
\label{sec:proposed}
Our proposed online SD system has two phases separated by a threshold $N_{init}$. 
Before the number of input embedding exceeds $N_{init}$, the system does not operate but only stacks the given embedding each time.
When the number of stacked embedding becomes greater than $N_{init}$, an initial clustering (see Section~\ref{ssec:init_clst}) is conducted. 
From then, our system outputs speaker labels for each input embedding (i.e., online phase, see Section~\ref{ssec:online_clst}).
Our online SD system adopts three hyper-parameters in total. $N_{init}$ decides when to start online operation, $N_{ckpt}$ corresponds to the maximum number of checkpoints used in the online phase, and distance threshold of centroid clustering (see Section~\ref{para:centroid_clustering}).

\subsection{Silhouette coefficient}
The silhouette coefficient~\cite{rousseeuw1987silhouettes} is a metric for analysing the reliability of clustering results. 
This metric measures an overall representative score, considering the cohesion of individual clusters and the separation between clusters.
We assume that the silhouette coefficient can indicate how well the clustering result models speakers within a session. 
Thus, we utilise the silhouette coefficient on various phases of our pipeline when estimating the number of speakers.

\subsection{Initial clustering phase}
\label{ssec:init_clst}
The system conducts clustering using the embeddings stacked so far when the number of embeddings exceeds $N_{init}$. 
Because initial clustering results heavily affect the result of the following online clustering phase, obtaining a better initial result is crucial. 

We usually tune the distance threshold of clustering to get a better result. However, we cannot adopt this strategy because we target the various data domains using only a single-setting system. Therefore, instead of tuning the threshold for each domain data, we adopt clustering with a silhouette coefficient trick. Some studies~\cite{kwon2021look,kwon2021adapting,kwon2022multi,kim2021disentangled} already composed their clustering-based SD systems using silhouette coefficient, and those systems show superior performance on various datasets without threshold tuning. 

We first calculate the silhouette coefficient varying the number of speakers, and the number of speakers with the highest score is selected for this step. We limit the maximum number of speakers in the initial clustering phase based on empirical observations to five. 
Then, the proposed dual buffers are initialised.
We initialise the checkpoint buffer with all embeddings input so far.\footnote{$N_{\textrm{init}}$ is identical to or smaller than $N_{\textrm{ckpt}}$.} 
Each centroid is initialised by averaging all embeddings corresponding to that speaker, leveraging the AHC result. 

\subsection{Online clustering phase}
\label{ssec:online_clst}
Figure~\ref{fig:framework} illustrates the overall scheme of our proposed online SD framework.

\newpara{Speaker number decision}
Once the online SD system starts to operate, every input embedding follows one of two processes based on whether it is judged as an embedding from an unseen speaker or a known speaker. 
We again adopt silhouette coefficients here to decide whether an input embedding is from an unseen speaker.
Specifically, we calculate the coefficient three times, configuring a different number of speakers and selecting the number of speakers where the coefficient value is the highest: (i) the number of current speakers - 1, (ii) the number of current speakers, and (iii) the number of current speakers + 1. 
We judge that a new speaker has appeared for case (iii). 
Case (i) exists based on our acknowledgement that an online SD system can falsely increase the number of speakers, which was empirically confirmed beneficial. 

\newpara{Unseen speaker.}
Once an input embedding is judged that it is from an unseen speaker, the system assigns a new global label to that speaker and outputs the label.
Then, we update the checkpoint by merging a pair of the most similar two checkpoint embeddings and add the newly input embedding. 
A new centroid is also initialised, where new input embedding is directly used.

\begin{figure}[t!]
  \centering
  \includegraphics[width=0.8\columnwidth]{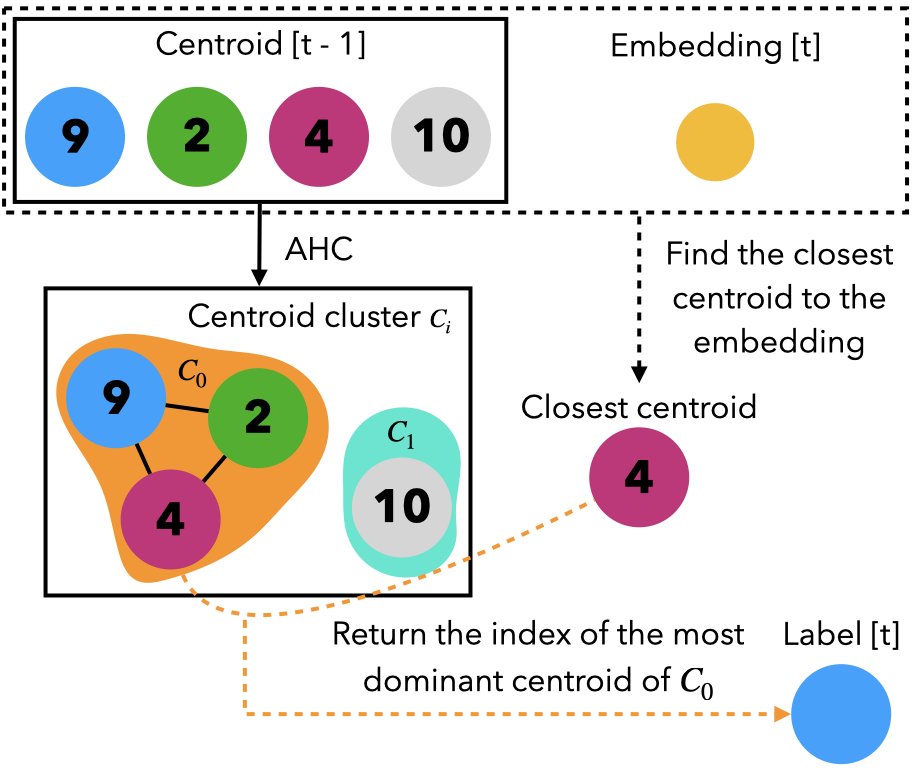}
  \caption{Label mapping with centroid clustering.}
  \label{fig:label_mapping}
\end{figure}

\newpara{Existing speaker: label mapping with centroid clustering}
\label{para:centroid_clustering}
When the input embedding belongs to an existing speaker, the system has to assign a global label without permutations.
For this process, we propose a method referred to as ``{\em centroid clustering}'', where we once again apply a clustering algorithm such as AHC to the centroids. 
This is based upon the philosophy mentioned in Section~\ref{sec:intro} that there may exist more than one centroid for a single speaker. 
Figure~\ref{fig:label_mapping} illustrates the process.
After the clustering is conducted, the online SD system finds the closest centroid in terms of cosine similarity (e.g., purple centroid in Figure~\ref{fig:label_mapping}. 
Then, it checks which centroid has been used for the most time within the cluster it belongs to (e.g., orange cluster) and outputs the label of that centroid (e.g., blue centroid, not purple). 
Thus, the output label can coincide with the closest centroid, but it can also be that of another centroid which belongs to the same cluster.

\section{Experiments}
\label{sec:exp}

\subsection{Datasets and metrics}
We evaluate our online SD system on four datasets: AMI Mix-Headset~\cite{mccowan2005ami,bredin2020pyannote}, test sets of the second and third DIHARD challenges~\cite{Ryant2019TheBaselines,ryant2020third}, and the test set of the VoxConverse~\cite{Chung2020SpotWild}.
AMI comprises meeting conversations.
The two DIHARD datasets involve several domains, including audiobooks and interviews.
VoxConverse includes diverse multimedia domains crawled from YouTube. 

We use the diarisation error rate (DER)  as the primary metric. 
DER consists of three components: false alarm (FA,  speech in prediction, but not in reference), missed speech (MS, speech in reference, but not in prediction), and speaker confusion (SC, speech which is assigned to the wrong speaker). 
DERs are measured using  {\tt dscore}\footnote{https://github.com/nryant/dscore}. 
We do not apply forgiveness collars for evaluation, except for VoxConverse, where we give 0.25s to enable comparisons with existing works.
We conduct all experiments using the oracle end point information since this paper aims to improve speaker confusion in real-time.

\input{tables/main_table}

\subsection{Implementation details}
\label{ssec:impl_details}
\newpara{Embedding extraction.} We first split the input audio into segments leveraging oracle end points.
Then we extract speaker embeddings using a sliding window with a $1.5$s window and a $0.5$s shift. 
We utilise the H / ASP architecture~\cite{kwon2020ins} as our model and prepare the model under the training protocol described in~\cite{kwon2021adapting}.

\newpara{Centroid clustering}
We adopt AHC as the centroid clustering algorithm of section~\ref{ssec:online_clst}.
Cosine distance is used for the clustering, and the linkage threshold of AHC is set to $0.25$ empirically for all experiments.\footnote{Although we have future plans to adopt silhouette coefficient here as well, at the current state, we could find a global threshold valid for all four datasets.}

\newpara{Hyper-parameters.}
Our system involves two hyper-parameters: $N_{init}$ and $N_{ckpt}$. 
These two parameters have been tuned for each dataset through a grid search. 
$N_{ckpt}$ is $180$ for all datasets. 
$N_{init}$ is set to $60$ for AMI and DIHARD, {\Romannumeral 3} and $120$ for DIHARD {\Romannumeral 2} and VoxConverse.

\section{Results analysis}
\subsection{Offline vs Online}
In Table~\ref{tab:main_table}, 
we compare the proposed online SD system with two baseline offline clustering-based SD systems.
All three systems share the identical speaker embedding extractor, and two offline baselines share the clustering algorithm addressed in~\ref{ssec:init_clst}.
The difference between offline-base and online is the clustering algorithm; offline-best refers to the system where we apply feature enhancement techniques proposed in \cite{kwon2021adapting}.

We observe surprisingly superior performance in DIHARD {\Romannumeral 2} and {\Romannumeral 3}, where our online SD system outperforms offline-base on both datasets and even offline-best in DIHARD {\Romannumeral 3} with a small margin.
In contrast, our system had 31.33\% and 36.81\% performance degradation compared to the offline-base in AMI and VoxConverse.
Further analysis regarding the inconsistent performance gap compared to corresponding offline systems is required. 
At the current state, we analyse that our system generalises well across diverse domains (DIHARD) but is not competitive when the number of speakers is four or more. DIHARD datasets also include sessions with more than four speakers, but most sessions comprise less than four speakers.

\subsection{Ablation study}
\input{tables/ablation_study_ckpt_threshold}

Table~\ref{tab:ablation_study_threshold_ckpts} describes the ablation results on the two hyper-parameters, $N_{init}$ and $N_{ckpt}$ evaluated on AMI and DIHARD {\Romannumeral 2}. 
For both datasets, $N_{init}$ had the best performance when set to 60 or 90 and $N_{ckpt}$ showed the best performance when the value was set to 150 or higher. 
However, the two hyper-parameters both did not affect the performance significantly.
We thus conclude that our system is not sensitive to the two hyper-parameters.

\subsection{Comparison with state of the art}
\input{tables/DH2_comparison}
We further compare with other widely adopted baseline offline SD systems and recent state-of-the-art online SD systems on DIHARD {\Romannumeral 2} and {\Romannumeral 3} datasets.
We establish a new state-of-the-art performance in both datasets, where our system outperforms the previous state-of-the-art model by 6.49\% in DIHARD {\Romannumeral 2}.

\subsection{Real-time factor}
\input{tables/realtime_factor}
We measured the real-time factors using an Intel(R) Xeon(R) CPU E5-2630 v4 @ 2.20 GHz CPU and an NVIDIA P40 GPU.
Our system used GPU for speaker embedding extraction, and CPUs conducted the other processes.

As shown in Table~\ref{tab:realtime_factor}, all of our pipelines' settings are enough to use in the real-time inference. In particular, for the most light-weighted setting, $N_{ckpt} = 60$, the proposed system can assign a speaker label to a 1.5s segment after an average delay of 0.034s.
It is lightweight enough to add a system end point detector and other pre/post-processing methods to the process pipeline in the future.
\label{sec:result}

\section{Conclusion and future works}
\label{sec:conclusion}
In this study, we made several design choices with the philosophy that an online SD system can be faulty in several aspects owing to its online nature. 
The resulting proposed online SD system starts operating in real-time after a certain period where it only stacks the input speaker embeddings fed every $0.5$ second.
The system also includes a mechanism which can decrease the estimated number of speakers, assuming that a new speaker found in the previous step might be wrong, where silhouette coefficients have been adopted to estimate the number of speakers.
Moreover, we adopt dual buffers and cluster the centroids, representing each speaker with the hypothesise that more than one centroid could have been generated for a single speaker.
Our system is state of the art in DIHARD {\Romannumeral 2} and {\Romannumeral 3} and demonstrates competitive performance compared to offline systems.

As the online SD is in its preliminary phase, and since our system is also new, several aspects exist to be further explored, which we leave as future works.
To generalise even better to more datasets, other methods than adopting a hard threshold can be applied when clustering centroids.
Leveraging the lightweight proposed system, additional pre/post-processing algorithms adapted for the online SD can be developed.
Further analysis should be done in situations where more than $5$ speakers utter.

\clearpage
\bibliographystyle{IEEEbib}
\bibliography{shortstrings,refs}

\end{document}

%% file: tables/main_table.tex
\begin{table}[!t]
  \centering
  \small
  \caption{
  Comparison of the proposed online SD system with two offline systems. ``offline-base'' excludes all feature enhancement techniques that can only be applied to an offline system and ``offline-best'' corresponds to our best performing offline system. 
  Performance reported on four evaluation datasets: AMI, DIHARD {\Romannumeral 2}, DIHARD {\Romannumeral 3}, and VoxConverse. 0.25s collar is applied to VoxConverse to match the literature and no collars are applied elsewhere. False alarm is zero for all datasets because we adopt reference end point detection. MS: miss, SC: speaker confusion, JER: Jaccard error rate.
	}
	\begin{tabularx}{\columnwidth}{lYYYY}
    \Xhline{1pt}
	 \textbf{Configuration} & \textbf{DER} & \textbf{JER} & \textbf{MS} & \textbf{SC} \\ 
	 \Xhline{1pt}
	 \multicolumn{5} {c} {\bf{AMI} } \\
     \hline\hline
	 offline-base & 20.27 & 29.20 &  14.55 & 5.72 \\
	 offline-best & \textbf{19.73} & 22.83 & 14.55 & 5.18\\
	 online & 22.88 & 28.81 & 14.55 & 8.33 \\
    \Xhline{1pt}
    \multicolumn{5} {c} {\bf{DIHARD {\Romannumeral 2}} } \\
     \hline\hline
     offline-base & 22.83 & 46.95 & 9.69 & 13.14 \\
     offline-best &  21.78 & 44.22 & 9.69 & 12.09\\
     online & \textbf{21.6} & 45.88 & 9.69 & 11.91 \\
    \Xhline{1pt}
    \multicolumn{5} {c} {\bf{DIHARD {\Romannumeral 3}} } \\
     \hline\hline
     offline-base & 20.38 & 39.20 & 9.52 & 10.86 \\
     offline-best & \textbf{18.97} & 35.59 & 9.52 & 9.45\\
     online & 19.05 & 38.98 & 9.52 & 9.52 \\
    \Xhline{1pt}
    \multicolumn{5} {c} {\bf{VoxConverse} } \\
    \hline\hline
    offline-base & 9.10 & 46.08 & 1.60 & 7.50 \\
    offline-best & \textbf{5.85} & 36.82 & 1.60& 4.25\\
    online & 13.47 & 50.79 & 1.60 & 11.87 \\
    \Xhline{1pt}
	\end{tabularx}
	\label{tab:main_table}
\end{table}


%% file: tables/ablation_study_ckpt_threshold.tex
\begin{table}[!t]
	\centering
	\small
	\caption{
	Ablation study about hyper-parameter combinations of $N_{init}$ and $N_{ckpt}$. Each value of the table is the DER of AMI Mix-Headset dataset.
	}
	\begin{tabularx}{\columnwidth}{c|YYYYY}
    \Xhline{1pt}
	\multirow{2}{*}{ \backslashbox[1mm]{$N_{init}$}{$N_{ckpt}$}} & 90 & 120 & 150 & 180 & 210 \\  \cline{2-6} 
     & \multicolumn{5}{c}{\textbf{AMI}}\\
     \hline\hline
      30 & 24.78 & 24.17 & 24.8  & 23.62 & 24.96 \\
      60 & 23.8  & 23.91 & 24.03 & \textbf{22.88} & 23.6  \\
      90 & 24.52 & 23.51 & 23.83 & 22.97 & 24.13 \\
    \Xhline{1pt}
     & \multicolumn{5}{c}{\textbf{DIHARD II}}\\
     \hline\hline
      30 & 23.95 & 23.49 & 23.43  & 23.64 & 23.24 \\
      60 & 22.53  & 22.38 & 22.14 & 22.49 & \textbf{21.83}  \\
      90 & 22.65 & 22.15 & 21.89 & 22.08 & 22.11 \\
     \Xhline{1pt}
	\end{tabularx}
	\label{tab:ablation_study_threshold_ckpts}
\end{table}

%% file: tables/DH2_comparison.tex
\begin{table}[!t]
	\centering
	\small
	\caption{
	Comparing proposed system with previous researches on DIHARD {\Romannumeral 2} and DIHARD {\Romannumeral 3} datasets.
	}
	\begin{tabularx}{\columnwidth}{lYY}
    \Xhline{1pt}
	 System & Type & DER \\ 
	 \Xhline{1pt}
	 \multicolumn{3} {c} {\bf{DIHARD {\Romannumeral 2}} } \\
     \hline\hline
	 Challenge Baseline~\cite{Ryant2019TheBaselines} & offline & 26.0 \\
	 VBx~\cite{landini2022bayesian} & offline & 18.55 \\
	 \hline
	 UIS-RNN-SML~\cite{fini2020supervised} & online & 27.3 \\
	 Flex-STB~\cite{xue2021online} & online & 25.8 \\
	 Core Sample~\cite{yue2022online} & online & 23.1 \\
	 Ours & online & \textbf{21.6} \\
	 \Xhline{1pt}
	 \multicolumn{3} {c} {\bf{DIHARD {\Romannumeral 3}} } \\
     \hline\hline
     Challenge Baseline~\cite{ryant2020third}  & offline & 19.25 \\
	 \hline
	 ChkptAHC~\cite{zhang2022low} & online & 19.57 \\
	 Core Sample~\cite{yue2022online} & online & 19.3 \\
	 Ours & online & \textbf{19.05} \\
	 \Xhline{1pt}
	\end{tabularx}
	\label{tab:dh2_comparison}
\end{table}


%% file: tables/realtime_factor.tex
\begin{table}[!t]
	\centering
	\small
	\caption{
	Measure the real-time factor of the proposed pipeline using 10-minute samples from the DIHARD {\Romannumeral 2} dataset. For each setting, measure the realtime factor 10 times and take the mean and standard deviation as values.
	}
	\begin{tabularx}{0.7\columnwidth}{lY}
    \Xhline{1pt}
	 $N_{ckpt}$ & Real-time Factor \\ 
     \hline
     60 & 0.02266 $\pm$ 0.00089 \\
     120 & 0.02433 $\pm$ 0.00070 \\
     180 & 0.02726 $\pm$ 0.00100 \\
     240 & 0.03115 $\pm$ 0.00070 \\
     300 & 0.03354 $\pm$ 0.00112 \\
     \Xhline{1pt}
	\end{tabularx}
	\label{tab:realtime_factor}
\end{table}